\documentclass[pra,twocolumn,eqsecnum,aps,showpacs]{revtex4}
\usepackage{graphicx}
\newtheorem{theorem}{Theorem} 

\newcommand{\Gm}{\Gamma}

\newcommand{\al}{\alpha}
\newcommand{\bt}{\beta}

\newcommand{\dl}{\delta}

\newcommand{\half}{\frac{1}{2}}
\newcommand{\ie}{\textit{i.e.,}}
\newcommand{\etal}{\textit{et al }}
 
\begin{document}
\title{A measure of non-Gaussianity for quantum states}
\author{J. Solomon Ivan}
\email{solomon@imsc.res.in}
\affiliation{The Institute of Mathematical Sciences, Tharamani, Chennai 
600 113}
\author{M. Sanjay Kumar}
\email{sanjay@bose.res.in}
\affiliation{S.N. Bose National Centre for Basic Sciences, 
Sector--III, Block--JD, Salt Lake, Kolkata 700 098}
\author{R. Simon}
\email{simon@imsc.res.in}
\affiliation{The Institute of Mathematical Sciences, 
C.I.T. Campus, Tharamani, Chennai 600 113}

\begin{abstract}
 We propose a measure of non-Gaussianity for quantum states 
of a system of $n$ oscillator modes. Our measure is based on 
the quasi-probability $Q(\alpha),\;\alpha\in{\cal C}^n$. 
 Since any measure of non-Gaussianity is necessarily an attempt 
at making a quantitative statement on the 
departure of the shape of the $Q$ function from Gaussian, 
 any good measure of non-Gaussianity 
should be invariant under transformations which do not alter the shape of 
the $Q$ functions, namely displacements, passage through 
passive linear systems, and uniform scaling of all the phase space 
variables: $Q(\alpha)\to \lambda^{2n}Q(\lambda \alpha)$. 
Our measure which meets this `shape criterion' 
 is computed for a few families of states, and the results are 
contrasted with existing measures of non-Gaussianity.
 The shape criterion implies, in particular, that the 
non-Gaussianity of the photon-added thermal states should 
be independent of  temperature. 

\end{abstract}
\pacs{03.67.-a, 03.65.-w, 02.50.Cw}
\maketitle

\section{Introduction}
Quantum information theory  of
continuous variable systems has been actively pursued in recent years,  
especially in the context of
Gaussian states\,\cite{cv-qi1,cv-qi2,cv-qi3}. 
 Such states are the ones which occur naturally in most 
 experimental situations, particularly in quantum optics. While these 
states
live  in an infinite-dimensional Hilbert space, they are  
 remarkably easy to
handle since they are fully described by their covariance matrix (and
first moments). Further, 
their evolution under quadratic Hamiltonians is easily cast in the 
language of symplectic groups and (classical) phase 
space\,\cite{simon00,simonbook, simon87}.  
The fundamental protocol of quantum teleportation  
has been achieved using these states\,\cite{bennett93, furusawa98}.
 However, there are situations wherein 
one  deals with (nonclassical) non-Gaussian 
resources to generate entanglement \cite{mary97, klyshko96, waks04,
waks06,solo06, asboth05,heersink06, molmer06}.  
They arise naturally in nonlinear evolutions like passage through a
Kerr medium\,\cite{schack90,tyc08}.

It has been shown recently that teleportation fidelities
can be improved with the use of 
non-Gaussian resources\,\cite{dellanno07}.
It is thus important that one 
is able to quantify the non-Gaussianity of such resources. 
 Effort in this direction 
has been initiated in some recent publications\,\cite{genoni07, 
genoni08}.

From the perspective of classical probability theory, Gaussian 
distributions are those probability distributions which
are completely specified by their first and second moments; all their
higher-order moments are determined by these lower-order
moments. 
Non-Gaussian probabilities do not enjoy this special property.
An easier, and possibly more effective, way to distinguish the two is 
through cumulants:
every non-vanishing cumulant of order greater
than two serves as an indicator of non-Gaussianity of the
probability distribution under consideration\,\cite{kendall,huber85}. 

The purpose of any good measure of non-Gaussianity in the context of 
classical probability theory
is thus to capture the essence of the non-vanishing higher-order cumulants.
A non-Gaussianity measure should thus manifestly depend on the
higher-order cumulants.
Yet another desirable feature one would like to 
have is  {\em invariance of the measure under scaling}.  
{\em Ultimately, non-Gaussianity measure is a quantitative statement 
of the departure of the shape of a probability distribution from 
Gaussian}.  But  {\em uniform scaling of \textit{all} 
 the variables of a 
probability distribution does not alter the  `shape' 
of the distribution, and hence it should not affect its 
non-Gaussianity}. 

The notion of non-Gaussianity can be extended 
to a quantum mechanical state through its definition on 
the associated $Q$ function, a member of the one-parameter family of 
$s$-ordered quasi-probabilities\,\cite{cahill691}. That this is 
an appropriate route is endorsed by the 
fact that the Marcinkiewicz theorem [see below] holds for the $s$-ordered 
quasi-probabilities as well. It turns out that the cumulants of order
greater than $2$ for the various 
$s$-ordered quasi-probabilities corresponding to a 
fixed state $\hat{\rho}$ 
are independent of $s$, 
indicating that the higher order cumulants are intrinsic to the state.  
Moreover,  
all higher-order cumulants of order greater than $2$ vanish 
identically for  Gaussian  states.
Thus any non-vanishing higher-order cumulant of the quasi-probability
indicates  non-Gaussianity of the state, and this conclusion is
independent of the ordering parameter $s$.

The above considerations will suggest  that any
good measure of non-Gaussianity relevant in the 
context of classical probability theory can, with suitable 
modification, lead to a  
good measure of non-Gaussianity of quantum mechanical states,  
 provided a state is identified 
through its $Q$ function 
(\,For a brief review on such measures in classical probability theory,
see \cite{huber85}\,).
 The purpose of such a quantum measure 
would be to capture the essence of the 
non-vanishing higher-order cumulants 
of the $Q$ function associated with the state.
 And invariance of the measure 
under an overall scaling of the $Q$ function 
 is a desirable feature worth insisting on.

The purpose of this paper is to motivate and present 
such a measure of non-Gaussianity of quantum states.  
Our measure is based on the Wehrl entropy\,\cite{wehrl79}, 
the quantum analogue 
of differential entropy\,\cite{cover} well-known from the 
context of classical information theory of 
continuous variables [Differential entropy itself 
is a generalisation of  Shannon entropy from discrete to 
continuous variables].  

The photon-added thermal states\,\cite{tara92} play a key 
role in our considerations. These nonclassical states 
have been generated 
experimentally\,\cite{zavetta07,parigi08,parigi07,kiesel08}. 
Their special importance to the present work arises from 
the fact that the $Q$ functions of these states are 
scaled versions of those of the Fock states, and therefore 
one will expect  any good measure of non-Gaussianity  to return 
the same values for both classes of states. 

The plan of the paper is as follows. In Section II we briefly
introduce the definition 
of moments and cumulants, and recall two well-known theorems 
in the context of these notions.
The one-parameter family of 
$s$-ordered quasi-probabilities corresponding to quantum density 
operators is briefly considered in Section III, 
with particular emphasis on the $Q$ function,  
and in  Section IV  we review the relationship between 
 differential entropy and the Kullback-Leibler distance  of classical 
probability theory. As a final item of preparation,
 we review briefly in Section IV the Wehrl entropy \cite{wehrl79} and 
some of
its properties. With these preparations, we introduce in Section VI 
 our non-Gaussianity measure  and explore some of its more important 
properties, including its invariance under uniform scaling of the 
underlying phase space.  
 In Section VII we evaluate this measure for three families of 
 quantum states, and in 
Section VIII we  compare our measure  
 with two  other measures of non-Gaussianity  
available in the literature. The paper concludes in Section IX 
with some additional remarks.
\section{Moments and cumulants}
For a multivariate probability distribution ${\cal P}(x)$, 
where $x=(x_{1},x_{2}, \cdots ,x_{n}) \in {\cal R}^{n}$, the
characteristic function $\chi(\xi),\;\xi\in {\cal R}^n$, is given by 
 the Fourier transform of ${\cal P}(x)$\,\cite{kendall}:
\begin{eqnarray*}
    \chi (\xi) &=& \int d^{n}x \, 
{\cal P}(x)\exp[\,{i\xi \cdot x}\,]           \nonumber \\
&=& \sum_{m_1 m_2 \cdots m_n}
\left(\prod_{k=1}^n\frac{{(i\xi_k)}^{m_k}}{m_k!}\right) 
\langle 
x_1^{m_1}x_2^{m_2} \cdots x_n^{m_n} \rangle \,,\nonumber
\end{eqnarray*}
\begin{eqnarray}
\langle x_1^{m_1}x_2^{m_2} \cdots x_n^{m_n} \rangle =
\int d^{n}x\, x_1^{m_1} x_2^{m_2} \cdots x_n^{m_n}
{\cal P}(x)\,.\;\, 
\label{ng1.1}
\end{eqnarray}
It follows from the invertibility of Fourier transformation that
the characteristic function retains all the information
contained in the probability distribution. The characteristic function
is often called the moment generating function, since one
obtains  from it all the moments of the underlying 
probability distribution through this compact expression:
\begin{eqnarray}
\langle x_1^{m_1}x_2^{m_2} \cdots x_n^{m_n} \rangle = 
\left(\prod_{k=1}^n 
\frac{d^{m_k}}{d{(i\xi_k)}^{m_k}}\right)
\chi (\xi)\, |_{\xi=0}\,. 
\label{ng1.2}
\end{eqnarray}

Another equivalent description of a probability distribution is
through  the cumulant generating function. This is 
defined through the logarithm of the characteristic function
\begin{eqnarray}
\Gamma (\xi) &=& 
{\rm log}\,\chi ( \xi) \nonumber \\
&=& \sum_{m_1 m_2 \cdots 
m_n}\left(\prod_{k=1}^n \frac{{(i\xi_k)}^{m_k}}{m_k!} 
\right){\gamma}_{m_1, m_2, \cdots, m_n}\,, \;\;
\label{ng1.3}
\end{eqnarray}
or, equivalently, through
\begin{equation}
\chi (\xi) = {\rm exp}(\Gamma(\xi))\,.
\label{ng1.4}
\end{equation}
From Eq. (\ref{ng1.3}), it is easy to see that
the cumulants  ${\gamma}_{m_1, m_2, \cdots, m_n}$ can be expressed as 
\begin{eqnarray}
{\gamma}_{m_1, m_2, \cdots, m_k} = 
\left(\prod_{k=1}^n 
\frac{d^{m_k}}{d{(i\xi_k)}^{m_k}}\right)
\Gamma (\xi)\, |_{\xi=0}\,. 
\label{ng1.5}
\end{eqnarray}
Thus, the cumulants are related to $\Gamma(\cdot)$ in 
precisely the same way as 
the moments are related to $\chi(\cdot)$. 
The set of all moments 
$\langle x_1^{m_1}x_2^{m_2} \cdots x_n^{m_n} \rangle$ gives a complete
characterisation of a probability distribution ${\cal P}(x)$, and the
same is true of the set of all cumulants ${\gamma}_{m_1, m_2, \cdots, 
m_n}$
as well. Indeed, one can describe one set in terms of the other
\cite{kendall, smith95,shohat-book}.

With these notations and definitions on hand, we now recall two 
important results from classical probability theory.

\begin{theorem}
The cumulant generating function of a Gaussian probability
distribution in $n$ variables is a multinomial of degree 
equal to 2\,\cite{kendall}.
\end{theorem}
\begin{theorem}
(Marcinkiewicz Theorem). If the cumulant generating function
of a (normalised) function in $n$ variables is a multinomial of finite 
degree
greater than 2, then the function will not be point wise non-negative,
and hence will fail to be a probability 
distribution\,\cite{marcinkiewicz39, rajagopal74}.
\label{marcinkiewicz}
\end{theorem}

Theorem 1 is a statement of the fact that a Gaussian probability 
 is fully determined by its moments of order $\le 2\,$; 
 all the higher-order 
cumulants are identically zero for a Gaussian probability.  
Theorem 2 is a much stronger statement.
It implies that any true probability distribution
other than the Gaussian distribution has a cumulant generating function
which cannot truncate at any (finite) order. That is, a non-Gaussian 
probability 
distribution has non-vanishing cumulants of arbitrarily high order.
 We note in passing that {\em non-vanishing cumulants of order greater 
than $2$ serve as indicators 
of the non-Gaussianity of the underlying probability}.

\section{Quasi-probabilities and the Q function}
A state of a quantum mechanical system specified 
by  density operator $\hat{\rho}$ can be faithfully described 
by any  member of the one-parameter family of 
$s$-ordered   quasi-probability distributions
 $-1\le s<1$\,\cite{cahill691}. 
In other words, an $s$-ordered quasi-probability captures all the 
information
present in the density operator $\hat{\rho}$. However, it is not 
a genuine probability distribution in general; in particular, it is not 
point wise non-negative. The prefix \textit{quasi} underscores 
precisely this aspect.
Nevertheless, the
$s$-ordered family of quasi-probability distributions
gives us a framework wherein one could give a phase space description 
of quantum mechanical systems  in the language of classical probability
theory. 

For a quantum state describing the radiation 
field of $n$ modes ($n$ oscillators) the characteristic function of the 
$s$-ordered quasi-probability, for any $-1\le s\le 1$, 
is defined through\,\cite{cahill691}
\begin{eqnarray}
    \chi_{\rho}(\xi; s) = \exp[\,\frac{s}{2} |\xi|^{2}\,]\, {\rm 
Tr}(\hat{\rho} 
D(\xi))\,,
\end{eqnarray}
where $\xi=(\xi_{1},\xi_{2}, \cdots ,\xi_{n}) \in {\cal C}^{n}$, and
$D(\xi)$ is the $n$-mode (phase space) displacement operator:
\begin{eqnarray}
    D(\xi) &=& \exp[\, 
\sum_{j}(\xi_{j}\hat{a}_{j}^{\dagger}-\xi^{*}\hat{a}_{j})\,]\,. 
\end{eqnarray}
The $s$-ordered quasi-probability itself is just the Fourier transform 
of this 
characteristic function $\chi_{\rho}(\xi; s)$:
\begin{eqnarray}
\!\!\!W_{\rho}(\al;s)\! = \!\! 
\int \!\!{\rm exp}
[\,\sum_{j} (\alpha_j^{*} {\xi}_{j} -{\alpha}_{j} \xi_j^{*})\,]
\chi_{\rho}(\xi; s)\prod_{j} d^2 \xi_j.
\end{eqnarray}
 Here  ${\hat{a}}_{j}$ and
${\hat{a}}_{j}^{\dagger}$ are the annihilation and creation operators
 of the $j$th mode, 
${\alpha}_{j}$ represents  the (c-number) phase space variables 
 $q_j,p_j$ corresponding to the $j$th
mode through ${\alpha}_{j} = (q_j + i p_j)/\sqrt{2}$,  
and $\alpha =(\alpha_1,\alpha_2, \cdots,\alpha_n) 
\in {\cal C}^n$.  The particular cases $s= -1, 0, 1$ correspond, 
respectively, to the better known 
$Q$ function, the Wigner function, and the $P$ function. 

The $Q$
function corresponding to a density operator $\hat{\rho}$ has a 
particularly simple expression in terms of coherent state projections:
\begin{eqnarray}
Q(\al) = \langle \al|\rho|\al \rangle,\;\;\; 
\alpha \in {\cal C}^n\,.
\end{eqnarray}
It may be noted that the $Q$ function 
is manifestly nonnegative for all 
$\alpha \in {\cal C}^n$.

Reality of $W_{\rho}(\al;s)$ is equivalent to  hermiticity of 
the density operator $\hat{\rho}$,  
and the fact that $\hat{\rho}$ is of unit trace faithfully transcribes to 
\begin{eqnarray}
\frac{1}{\pi^n}\,\int W_{\rho}(\al;s) d^2 \alpha= 1\,.
\end{eqnarray}
While these two properties hold for every $s$-ordered 
quasi-probability, 
point wise  non negativity for all states is a distinction 
which applies  to the $Q$ function alone. In other words, {\em the $Q$ 
function is a genuine probability distribution;
 every  other  $W_{\rho}(\al;s)$ is only a quasi-probability}.
Gaussian pure states are the only pure states for which the Wigner 
function is a  classical probability\,\cite{hudson}; in the case of 
 $P$ function, the coherent states are the only pure state with this 
property.

However, not every probability distribution is  a $Q$ function. 
This is evident, for instance, from the obvious fact that $Q(\alpha)\le 
1,\;\,\forall\,\alpha \in{\cal C}^n$. 

The next result captures, in a concise form, the manner in which  
 members of the one-parameter family of
$s$-ordered quasi-probabilities $W_{\rho}(\al;s)$ differ from one 
another for a given  state $\hat{\rho}$.
\begin{theorem}
Only the second order cumulants of the quasi-probability of a given 
state 
depend on the order parameter $s$; all the other cumulants are 
 independent of the quasi-probability under consideration.
\end{theorem}

This result is already familiar in the case of 
a single-mode radiation field \cite{schack90}. But the proof is, 
as outlined below,  
immediate in the multi-mode case as well.  
 The characteristic functions of a state $\hat{\rho}$ for two different
values of the order parameter $s_1$ and $s_2$ are 
obviously related in the following manner\,\cite{cahill691}:
\begin{eqnarray}
&&\chi_{\rho}(\xi ; s_1) =
{\rm exp}\left((s_1-s_2){|\xi|}^2\right) \chi_{\rho}(\xi; s_2)\,.
\end{eqnarray}
On taking logarithm of both sides to obtain the corresponding cumulant 
generating functions we have
\begin{eqnarray}
&&{\rm log}\chi_{\rho}(\xi; s_1) =
(s_1-s_2) {|\xi|}^2 + {\rm log}\chi_{\rho}(\xi; s_2)\,.
\nonumber
\end{eqnarray}
That is, 
\begin{eqnarray}
&&\Gm_{\rho}(\xi; s_1) =
(s_1-s_2) {|\xi|}^2 +\Gm_{\rho}(\xi ; s_2)\,.
\end{eqnarray}
Thus the cumulant generating function for different $s$-ordered 
quasi-probabilities differ only in  second order, completing proof of 
the theorem.  

In these equations $|\xi|^2$ stands, as usual, for 
$\sum_{j=1}^{n} {|\xi_j|}^2$. As an immediate consequence of this 
theorem we have

\begin{theorem}
 For no quasi-probability can the cumulant generating function 
 be a multinomial of finite order $> 2$.
\end{theorem}

\noindent
{\em Proof}\,: Since the $Q$ function, for every state $\rho$,  
is a genuine probability distribution, it follows from    
the Marcinkiewicz theorem that the cumulant generating function 
 of  $Q$ cannot be a multinomial of finite order $> 2$. 
Since the different $s$-ordered quasi-probabilities differ 
only in second-order cumulants, this conclusion holds for all 
$s$-ordered quasi-probabilities, thus proving the theorem.

We conclude this Section with the following remarks. The 
above considerations show that  
quasi-probabilities fail to be true probabilities only 
in this limited sense: they differ from  genuine probabilities  
only in cumulants of order two. The distributions, however,  
can be quite different from classical probabilities, particularly for 
$s>0$, and they can become  
as subtle as Fourier transform of $\exp\,[\,\sigma\,y^2\,], \;\sigma>0$, 
a Gaussian with the wrong signature for the variance. 

Since the higher-order cumulants, 
which should play an essential role in any reasonable definition of 
non-Gaussianity measure, do not depend on the value of the parameter 
$s$, they may be viewed as attributes intrinsic to the state under 
consideration; we may therefore use any convenient quasi-probability 
to capture their essence.

\section{Differential entropy and the Kullback-Leibler distance }
The role of Shannon entropy of probability distributions over discrete 
random variables is taken over by 
differential entropy in the case of continuous variables.  
 Given a multivariate probability distribution ${\cal P}(x)$ in $n$ 
variables $(x_1,x_2,\cdots,x_n)\in {\cal R}^n$, the associated 
differential entropy $H({\cal P}(x))$ is defined by\,\cite{cover} 
\begin{eqnarray}
 H({\cal P}(x))= -\int d^n x{\cal P}(x){\rm log}\, 
{\cal P}(x)\,.
\label{ng2.1}
\end{eqnarray}   
But unlike the Shannon entropy, the differential entropy can be 
negative.
This is manifest, for instance, for uniform distribution over a region 
  of less than unit volume in ${\cal R}^n$.

Among all the probability distributions with a fixed set of first and 
second
moments, the Gaussian probability distribution has the maximum 
differential entropy \,\cite{cover}. This fact may be used to 
 modify  differential entropy to result in a non-negative quantity
\begin{eqnarray}
 J({\cal P}(x)) = H({\cal P}_G (x)) - H({\cal P}(x))\,.
\label{ng2.2}
\end{eqnarray} 
Here ${\cal P}_G (x)$ is the Gaussian probability distribution
\textit{with the same first and second moments} as the given 
probability
distribution ${\cal P}(x)$. 

It may be recalled that Kullback-Leibler distance between two 
probabilities $P_{1}(x)$
and $P_{2}(x)$ is defined as the difference of their 
 differential  entropies \cite{cover}: 
\begin{eqnarray}
    S({\cal P}_{1}(x)||{\cal P}_2(x))
    &=& H({\cal P}_2(x)) - H({\cal P}_{1}(x)) \nonumber \\
    &=& -\int {\cal P}_{1}(x){\rm log} 
    ({\cal P}_{1}(x)) d^n x\nonumber\\
  &&~ +
    \int {\cal P}_{2}(x){\rm log} 
    ({\cal P}_{2}(x))  d^n x\, .\;\;\;\;
\label{ng2.3}
\end{eqnarray} 
Thus  $J({\cal P}(x))$  can be regarded as the
Kullback-Leibler distance 
between the given probability $P(x)$ and
 the associated Gaussian distribution $P_{G}(x)$:
\begin{eqnarray}
    J({\cal P}(x)) = S({\cal P}_G(x)||{\cal P}(x))\,.
\end{eqnarray}
    $J({\cal P}(x))$ is sometimes known by the name negentropy.

\section{Wehrl entropy}
Wehrl entropy\,\cite{wehrl79,wehrl78}  may be viewed as 
the extension of differential entropy to the quantum mechanical 
context, but the Wehrl entropy has interesting properties which 
distinguish it from
differential entropy. The distinction arises from
the fact that while every  $Q$ function certainly qualifies to be  a 
classical probability distribution,
every classical probability is not a $Q$ function.
The uncertainty principle has a fundamental role to
play in this aspect\,\cite{wehrl79}. 
The potential use of Wehrl entropy 
as a measure of the `coherent' component of a state has been  discussed 
in Ref\,\cite{orlowski93}. 
And its possible role in defining an entanglement
measure has also been explored\,\cite{mintert04, marchiolli08}.

 For a state $\hat{\rho}$ describing $n$ modes
of radiation field, the Wehrl entropy  is defined as
\begin{eqnarray}
    H_{W}(\hat{\rho})&=& - \frac{1}{\pi^{n}} \, \int \, \prod d^2 \alpha_j 
    Q_{\rho}(\alpha) {\rm log}\, Q_{\rho}(\alpha)\,,
\end{eqnarray} 
where $Q_{\rho}(\al)$ is the  $Q$ function corresponding to $\hat{\rho}$.  
 This definition may be compared with that of differential entropy;  
the role of ${\cal P}(x)$ in differential entropy 
is played by $Q_\rho(\alpha)$ in Wehrl entropy. 

However, in contradistinction  to differential entropy, the Wehrl 
entropy is always positive. This is an immediate consequence of the 
fact that the Q function is bounded from above by unity.  It turns out 
that the Wehrl entropy is always greater than or equal to 
unity\,\cite{lieb78};  
 indeed, it attains its least value of
unity for the coherent states and only for these states. 
This property can be thought of 
as a manifestation of the uncertainty principle, 
which the coherent states saturate. 
 Further, the Wehrl entropy  
is always greater than the von Neumann entropy\,\cite{wehrl79}: 
\begin{eqnarray}
 H_{W}(\hat{\rho}) \geq S(\hat{\rho})  =-\rm{Tr}(\hat{\rho}\, 
log\,\hat{\rho})\,.
\end{eqnarray}
While the von Neumann entropy is zero for pure states, we have just 
noted that the Wehrl 
    entropy $H_{W}(\hat{\rho})$ is greater than or equal to 
    unity for all states. 
    Several aspects of the Wehrl entropy have been  
    explored in Ref.\,\cite{orlowski93}.

\section{A non-Gaussianity measure for quantum states}

As is well-known, a
quantum state $\hat{\rho}$ is said to be Gaussian iff 
the associated Wigner distribution is Gaussian.
 This will suggest that the non-Gaussianity of a state is  coded into  
 the non-vanishing cumulants of order $>2$ of the Wigner function. 
 Since the Wigner and $Q$ functions are related by 
convolution by a Gaussian, 
 the $Q$ function of a state is Gaussian iff the 
Wigner function is, and the non-Gaussianity should thus 
be found coded in the 
higher-order cumulants of the $Q$ function as well. The consistency of 
these statements is ensured by the fact that the higher-order 
cumulants are the same for the Wigner and the $Q$ functions 
 [Indeed, as we have shown earlier, the
higher-order cumulants are intrinsic to the state, and hence  
 are the same for all $s$-ordered quasi-probabilities].

Non-Gaussianity can thus be described using 
either the Wigner function or the $Q$ function. 
 The fact that the $Q$ function is everywhere 
non-negative, rendering it a genuine probability 
in the  classical sense, makes it our preferred 
choice. We employ therefore the Wehrl entropy to capture the 
essence of the higher-order cumulants.

Given a state $\hat{\rho}$, our measure of non-Gaussianity 
${\cal N}(\hat{\rho})$ is defined as the difference of two Wehrl
 entropies:
\begin{eqnarray}
   {\cal N}(\hat{\rho})&=& H_{W}(\hat{\rho}_{G})-H_{W}(\hat{\rho})\,. 
\label{ng2.4}
\end{eqnarray}
Here $H_{W}(\hat{\rho})$ is the Wehrl entropy of 
the given state $\hat{\rho}$ and
$H_{W}(\hat{\rho}_{G})$ is the Wehrl entropy of the Gaussian state
$\hat{\rho}_{G}$ that has the same first and second moments as 
 $\hat{\rho}$. Since ${\cal N}(\hat{\rho})$ measures the 
departure of the Wehrl entropy of $\hat{\rho}$ from that of 
its Gaussian partner $\hat{\rho}_{G}$, it can be viewed as 
 a quantum Kullback-Leibler distance.  ${\cal N}(\hat{\rho})$ 
could also be viewed as a relative Wehrl entropy. 
But we prefer to call it simply a non-Gaussianity measure. 

This measure of non-Gaussianity enjoys 
several interesting properties. We will now list some of them:

(i) ${\cal N}(\hat{\rho}) \ge 0$,  equality holding
iff $\hat{\rho}$ is Gaussian.
 
\noindent
{\em Proof}\,: This is a restatement of the fact that 
the Wehrl entropy of a Gaussian state is greater than that 
of all states with the same first and second moments as the Gaussian.

(ii) ${\cal N}(\hat{\rho})$ is invariant under phase space displacements:
\begin{eqnarray}
 {\cal N}(\hat{\rho})={\cal N}(\,D(\xi)\, \hat{\rho}\, D(\xi)^{\dagger}\,)\,.
\end{eqnarray} 

\noindent
{\em Proof}\,: Let $D(\xi)\, \hat{\rho}\, D(\xi)^{\dagger}$ 
be denoted, for brevity, by $\hat{\rho}^{\,\prime}$. The $Q$ function of 
$\hat{\rho}^{\,\prime}$ is related to that of  $\hat{\rho}$ in this 
simple 
manner:
\begin{eqnarray}
Q_{\hat{\rho}^{\,\prime}}(\alpha)  = Q_{\hat{\rho}}(\alpha -\xi)\,.     
\end{eqnarray}
That is, displacement $D(\xi)$ acts as a rigid translation in phase 
space\,\cite{wehrl79,lieb78,dutta94}. Thus it has no effect  
on the Wehrl entropy of any state, and hence leaves ${\cal N}(\hat{\rho})$ 
invariant for every state.

(iii) ${\cal N}(\hat{\rho})$ is invariant under passage through 
any passive linear  system.

\noindent
{\em Proof\,}: A passive linear system is represented by a 
$n\times n$ unitary matrix $U$. It maps a coherent state 
$|\alpha\rangle$ into a new coherent state $|\alpha'\rangle 
= |U\,\alpha\rangle$\,\cite{wehrl79,lieb78,dutta94}, 
 where $\alpha \in {\cal C}^n$ is to be viewed as a 
column vector. Let $\hat{{\cal U}}_U$ be the unitary operator 
in the $n$-mode Hilbert space which represents 
the passive linear system labelled by the matrix $U$.
Let us denote by $\hat{\rho}^{\,\prime}$ 
the transformed state 
$\hat{{\cal U}}_U\,\hat{\rho}\,\hat{{\cal U}}_U^{\,\dagger}$ 
at the output of this passive system. Then the output  
$Q$ function is related to the input $Q$ function in this manner:
\begin{eqnarray}
Q_{\hat{\rho}'}(\alpha)  = 
Q_{\hat{\rho}}(U^{-1}\alpha) =
Q_{\hat{\rho}}(U^{\dagger}\alpha)\,.     
\end{eqnarray}
That is, the action of a passive linear system 
is a rigid $SO(2n)$ rotation in the $2n$-dimensional phase space.  
It follows immediately that this transformation does not change 
the Wehrl entropy of any state, and hence does not affect 
${\cal N}(\hat{\rho})$.

\noindent
{\em Remark}\,: While in the single-mode case of 
two-dimensional phase space all proper rotations 
are canonical transformations, this is not true in the 
multi-mode case. 
That is, $\rm{Sp}(2n, {\cal R})
\cap \rm{SO}(2n)$ is a proper subgroup of 
$\rm{SO}(2n)$ isomorphic to $\rm{U}(n)$, 
the $n^2$-parameter group of $n\times n$ unitary matrices, whereas 
$\rm{SO}(2n)$ is a 
much larger $(2n^2-n)$-parameter group\,\cite{dutta94}. 
 Only those phase space rotations which are 
elements of  this intersection 
act as unitary transformations in the Hilbert space of $n$ oscillators.
 
(iv) ${\cal N}(\hat{\rho})$ is invariant under a uniform phase space scaling
$\lambda$  defined at the level of  the $Q$ function in the following 
manner:  
\begin{eqnarray}
  \lambda\,:\;\;\;  Q(\alpha) \to Q'(\alpha)=\lambda^{2n}Q(\lambda 
\alpha)\,.
\end{eqnarray}

\noindent 
{\em Proof}\,: Under this uniform phase space scaling of the $Q$ 
function,  
the Wehrl entropy changes by a simple additive part {\em that is
independent of the state}:
\begin{eqnarray}
    H_{W}(\hat{\rho}) &=& -\frac{1}{\pi^n} \int \,Q(\alpha) 
\textrm{log}\,Q(\alpha)\prod_{j=1}^n d^2\alpha_j
          \nonumber \\
    & \to & -\frac{1}{\pi^n} \int 
    \lambda^{2n}Q(\lambda \alpha) \textrm{log}\,(\,\lambda^{2n}Q(\lambda 
\alpha)\,)\prod_{j=1}^n d^2\alpha_j 
\nonumber \\
    &  = & H_{W}(\hat{\rho}) -2n\, \rm{log}\,\lambda\,.
\label{ng25}
\end{eqnarray}
Note that in arriving at the last equation we have made a change of
variables in the integral and made use of the normalisation of the
$Q$ function. Now it trivially follows from this result that
${\cal N}(\hat{\rho})$, being a difference of two Wehrl entropies, remains
invariant.

\noindent
{\em Remark}\,: While the above conclusion holds 
mathematically for all $\lambda >0$, the scaled $Q$ function 
fails  to be a physical $Q$ function if $\lambda>1$. 
Therefore we restrict this scale parameter 
to the physically relevant range $0 <\lambda\le1$.

(v) ${\cal N}(\hat{\rho})$ is additive on tensor product states:
\begin{eqnarray}
{\cal N}({\hat{\rho}}_1 \otimes {\hat{\rho}}_2) = {\cal N}({\hat{\rho}}_1) + 
{\cal N}({\hat{\rho}}_2)\,.
\label{ng2.7}
\end{eqnarray}

\noindent
{\em Proof}\,: Under tensor product the $Q$ functions go as product
probabilities by definition. This is true of their associated 
Gaussian probabilities as well.

\noindent
{\em Corollary}\,:  For a bipartite state of the form 
$\hat{\rho} = {\hat{\rho}}_{a} \otimes {\hat{\rho}}_{G}$, where
${\hat{\rho}}_{G}$ is a Gaussian state
\begin{eqnarray}
{\cal N}(\hat{\rho}) = 
{\cal N}({\hat{\rho}}_{a} \otimes {\hat{\rho}}_{G})=
  {\cal N}({\hat{\rho}}_{a})\,.
\label{ng2.8}
\end{eqnarray}

\noindent
{\em Proof}\,: From (v) we have
\begin{eqnarray}
&&{\cal N}(\hat{\rho}) = {\cal N}({\hat{\rho}}_{a} \otimes {\hat{\rho}}_{G}) =
{\cal N}({\hat{\rho}}_{a}) + {\cal N}({\hat{\rho}}_{G})\,. \nonumber
\label{ng2.9}
\end{eqnarray}
and from (i)
\begin{eqnarray}
{\cal N}({\hat{\rho}}_{a}) + {\cal N}({\hat{\rho}}_{G}) = 
{\cal N}({\hat{\rho}}_{a})\,.
\label{ng2.10}
\end{eqnarray}

\noindent
{\em Proposition:} For a bipartite state of the form ${\hat{\rho}}_{\rm 
out}
= \hat{{\cal U}}_{U}\, ({\hat{\rho}}_{a} \otimes | \alpha \rangle \langle 
\alpha|) 
\,\hat{{\cal U}}_{U}^{\,\dagger}$, where $U$ represents a passive linear 
system  
and $| \alpha \rangle$ is a coherent state, we have 
\begin{eqnarray}
&&{\cal N}({\hat{\rho}}_{\rm out}) = {\cal N}({\hat{\rho}}_{a})\,.
\label{ng2.11}
\end{eqnarray}

\noindent 
{\em Proof}\,: From (iii) we have
\begin{eqnarray}
{\cal N}({\hat{\rho}}_{\rm out}) = 
{\cal N} ({\cal U}_{U}\, ({\hat{\rho}}_{a} \otimes | \alpha \rangle \langle 
\alpha|) 
\,{\cal U}_{U}^{\dagger}) =
{\cal N}({\hat{\rho}}_{a} \otimes |\alpha \rangle \langle \alpha|)\,.\;\; 
\nonumber
\label{ng2.12}
\end{eqnarray}
We have from (v)  
\begin{eqnarray}
&&{\cal N}({\hat{\rho}}_{a} \otimes | \alpha \rangle \langle \alpha|) =
{\cal N}({\hat{\rho}}_{a}) + {\cal N}(| \alpha \rangle \langle \alpha|)\,. \nonumber
\label{ng2.13}
\end{eqnarray}
Since the coherent state $|\alpha\rangle$ is Gaussian, we have from (i) 
\begin{eqnarray}
&&{\cal N}({\hat{\rho}}_{a}) + {\cal N}(| \alpha \rangle \langle \alpha|)=
{\cal N}({\hat{\rho}}_{a})\,.
\label{ng2.14}
\end{eqnarray}

This result  is useful in evaluating the  
non-Gaussianity of bipartite states 
 produced by the action of beam splitters, as we shall 
illustrate in the next Section.

\subsection{Shape criterion for good measure of non-Gaussianity}

Properties (ii), (iii), and (iv) deal with transformations 
which do not change the shape of the $Q$ functions. 
Since non-Gaussianity is a quantitative statement 
regarding the departure of the shape of the $Q$ function 
from Gaussian, it will appear that any good measure of 
non-Gaussianity should return the same value for all states 
connected by these transformations.  
In particular, two quantum states whose
$Q$ functions are related by a uniform scaling of
all the phase space coordinates should be assigned the
same amount of non-Gaussianity.
 We will call this the {\em shape criterion}, and 
we have seen that our measure 
${\cal N}(\hat{\rho})$ meets this requirement. 

\section{Examples}
In this Section we evaluate our non-Gaussianity measure ${\cal N}(\hat{\rho})$ 
for three families of states, namely the Fock states, the 
photon-added 
thermal states, and the phase-averaged coherent states of a single-mode 
of radiation. While the first two families consist of nonclassical 
states, the third one is a family of classical states. 

\subsection{Photon number states}

The $Q$ function of the Fock state (energy eigenstate) 
 $\hat{\rho}=|m\rangle \langle m |$ of the oscillator  
is given by the phase space distribution 
\begin{eqnarray}
    Q_{|m\rangle}(\alpha)= \frac{{|\alpha|}^{2m}}{m!} \exp ({-{|\alpha|}^2})\,,
\end{eqnarray}
whose only non-vanishing moment of order $\leq 2$  is
$\langle {|\alpha|}^2 \rangle= \textrm{Tr}(\hat{\rho} \hat{a}\hat{a}^{\dagger}) 
= m+1$.
 The phase space average $\langle {|\alpha|}^2 \rangle$ is with respect to 
the probability distribution $Q_{|m\rangle}(\alpha)$ and, by definition, it 
equals the (quantum) expectation value of the associated anti-normally  
ordered operator  $\hat{a}\hat{a}^{\dagger}$.
 The Gaussian state
which has the same moments of order $\leq 2$ as  
$\hat{\rho}_{|m\rangle} =
|m \rangle \langle m|$ is clearly  the thermal state with mean photon 
number $\langle \hat{n} \rangle \equiv \langle\hat{a}^{\dagger}\hat{a} 
\rangle = m$. The $Q$ function of 
such a thermal state $\hat{\rho}_{G}$ is given by
\begin{eqnarray}
Q_{G}(\alpha) = \frac{1}{\langle \hat{n} \rangle +1}
               \exp \left({-\frac{{|\alpha|}^2}{\langle 
\hat{n} \rangle +1}}\right), \,\,\, \langle \hat{n} \rangle = m\,.
\end{eqnarray}

The Wehrl entropy corresponding to $\rho_{G}$ is easily computed:
\begin{eqnarray}
    H_{W}(\hat{\rho}_{G}) & = & 1+\textrm{log}(1+\langle \hat{n} \rangle) 
\nonumber \\
    & = & 1+\textrm{log}(1+m)\,.
\label{ng2.15}
\end{eqnarray}
The Wehrl entropy of the photon number 
state $\hat{\rho}=|m\rangle \langle m |$  is 
\begin{eqnarray}
H_{W}(\hat{\rho}_{|m\rangle}) = -\frac{1}{\pi} \int d^2 \alpha Q_{| m \rangle}
(\alpha) {\rm log} Q_{| m \rangle}(\alpha)\,.
\end{eqnarray}  
This can be computed explicitly
by going to the polar coordinates, and one obtains\,\cite{orlowski93}
\begin{eqnarray}
H_{W}(\hat{\rho}_{|m\rangle}) &=& 1+m+{\rm log}m! -m \psi(m+1), \nonumber\\
&&\psi(m+1) = \sum_{k=1}^{m} \frac{1}{k} -\gamma\,,
\end{eqnarray}
where $\psi(m)$ is the digamma function,
 and $\gamma=0.5772\cdots$ is 
the Euler constant.
Hence the non-Gaussianity of the photon number state
$\hat{\rho}=|m\rangle \langle m |$ is 
\begin{eqnarray}
\!{\cal N}(\hat{\rho}_{|m \rangle})\! &=&\! 
{H}_{W}(\hat{\rho}_{G}) -H_{W}(\hat{\rho}_{|m\rangle}), 
\nonumber \\ 
\!\!&=& \!{\rm ln} (m+1) - m - {\rm log}m! + m \psi(m+1).\;\;\;
\end{eqnarray}

\begin{figure}[htb]
\begin{center}
\scalebox{0.7}{\includegraphics[width=10cm]{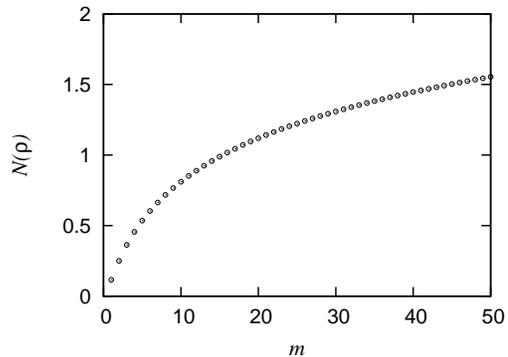}}
\caption{Variation of ${\cal N}(\rho)$ with number of photons $m$ 
for the Fock state $\rho=|m\rangle \langle m|$.
\label{ngfig1} }
\end{center}
\end{figure}

In Fig.(\ref{ngfig1}) we have plotted this non-Gaussianity as a 
function
of the photon number $m$. It is clear that the non-Gaussianity 
of $|m\rangle$ increases
monotonically with the photon number $m$, and goes to $\infty$ as $m$
tends to $\infty$. That this was to be expected can be  seen as 
follows.  For large
$m$ values $\psi(m+1) \sim \textrm{ln}(m+1)$, and $\textrm{log}m! \sim
m\textrm{log}m - m$, and hence ${\cal N}(\hat{\rho}_{|m \rangle}) 
\sim \textrm{log}(m+1)$. 
We shall be returning to this result in the next Section.

 Now consider a bipartite state of two modes with one mode in the Fock 
state and the other in the vacuum. Non-Gaussianity 
of this product  state is the same as 
that of the Fock state, and  this follows from Eq. (\ref{ng2.8}) . 
Let this bipartite state be passed through a beam splitter. 
 The state at the output will be entangled due to the 
nonclassicality of the Fock state\,\cite{asboth05,solo06}, 
but in view of Eq. (\ref{ng2.11}), this two-mode state will have the same 
non-Gaussianity as the original single-mode Fock state.

\subsection{Photon-added thermal states}
In this subsection we evaluate the non-Gaussianity of the photon-added
thermal state (PATS)\,\cite{tara92}. 
 The PATS is defined through 
\begin{eqnarray}
\hat{\rho} = C\, {\hat{a}}^{\dagger m}
{\hat{\rho}}_{\rm th} {\hat{a}}^{m}\,,
\label{ng3.1}
\end{eqnarray}
where $C$ is the normalisation constant which ensures ${\rm 
Tr}\,(\hat{\rho}) = 1$, and ${\hat{\rho}}_{\rm th}$ is
the thermal state given by
\begin{eqnarray}
{\hat{\rho}}_{\rm th} = (1-x)\sum_{n=0}^{\infty} x^k 
| k \rangle \langle k |\,;\;\; 
x=\exp\left[\,{-\frac{\hbar \omega}{kT}}\,\right]\,.
\label{ng3.2}
\end{eqnarray}
One can alternatively define the PATS through
parametric differentiation:
\begin{eqnarray}
\hat{\rho} = \frac{{(1-x)}^{m+1}}{m!} 
\frac{d^m}{dx^m}\sum_{k=0}^{\infty} x^k | k \rangle \langle k|\,.
\label{ng3.3}
\end{eqnarray} 
PATS are thus parametrised by two parameters: $0\le x<1$, and 
$m= 0,1,2,\cdots\;$. The limit $x\to 0$  corresponds to Fock states, 
and the limit  $m\to 0$ corresponds to thermal states.

We may note that PATS (with $m \ge 1$) is nonclassical for all values 
of $x$\,\cite{solo06}. Indeed, it violates a three-term classicality 
condition\,\cite{mary97}.  

The $Q$ function of PATS can be easily calculated and is given by
\begin{eqnarray}
Q_{\rm PATS}^{(m,x)}(\alpha) = \frac{{(1-x)}^{m+1}}{m!} {|\alpha|}^{2m} {\rm 
exp}[-(1-x){|\alpha|}^{2}]\,.\;\;
\label{ng3.4}
\end{eqnarray} 
It is evident that {\em the $Q$ function of the PATS is
a scaled version of the $Q$ function of the Fock state}:
\begin{eqnarray}
    Q_{\rm PATS}^{(m,x)}(\alpha)
= \lambda^2 Q_{| m \rangle}(\lambda\, \alpha), 
\;\;\lambda = \sqrt{1-x\,}\,.
\label{ng3.5}
\end{eqnarray}
Since our measure of non-Gaussianity respects the shape criterion
put forward in the previous Section, it is immediate that
the non-Gaussianity of the PATS is the same as
that of the photon number state:
\begin{eqnarray}
   N(\,\hat{\rho}_{\rm PATS}^{(m,x)}\,)
&=& {\rm ln} (m+1) - m - {\rm log}m! + m \psi(m+1)\nonumber\\
&=& N(\hat{\rho}_{|m\rangle})\,.
\end{eqnarray}

It is worth emphasising here that the PATS is a special state with
regard to the question of
 verifying whether a given measure of non-Gaussianity is a good
measure, \ie \,whether it satisfies the shape criterion. The test is
as simple as checking whether the measure in question evaluated
for the PATS is independent of the temperature parameter $x$ or not.
\begin{figure}[htb]
\begin{center}
\scalebox{0.7}{\includegraphics[width=10cm]{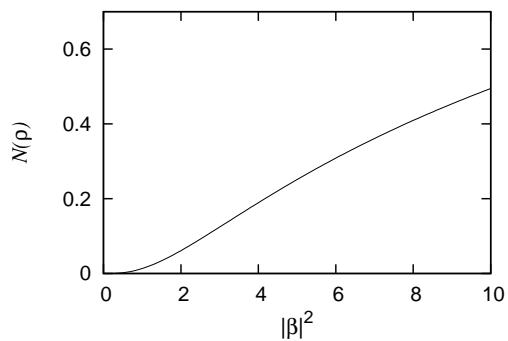}}
\caption{Variation of ${\cal N}(\rho)$ with energy $|\bt^{2}|$ for
the phase-averaged coherent state.
\label{ngfig2}}
\end{center}
\end{figure}

Finally we consider, as in the previous Subsection, a bipartite state of 
two modes, with one mode in the PATS   ${\hat{\rho}}_{\rm PATS}^{(m,x)}$ 
and the other in the vacuum state. 
Let us pass this two-mode state  through a beam splitter. 
That the state at the output of the beam splitter is entangled follows 
from the nonclassicality of the PATS\,\cite{solo06, asboth05}. 
It follows from Eq. (\ref{ng2.11}) 
that  non-Gaussianity of this entangled 
state is the same as that of the PATS, and hence is fully determined by 
$m$. 

We have already noted that PATS violates a three-term classicality 
condition. This implies that the output state is entangled and, moreover,  
that it is single-copy distillable\,\cite{solo06}.

\subsection{Phase-averaged coherent states}

As our final example, we evaluate the non-Gaussianity for the
phase-averaged coherent states. 
 Given a coherent state $|\beta \rangle$ its phase-averaged version is
\begin{eqnarray}
{\hat{\rho}}_{|\beta|} &\equiv & \int \frac{d\theta}{2\pi} \, 
\exp [\,-i\,\theta\,\hat{a}^\dagger \hat{a}\,]\,|\beta\rangle \langle \beta |\,
\exp [\,i\,\theta\,\hat{a}^\dagger \hat{a}\,]\nonumber\\
 &=&{\rm exp}({-{|\beta|}^2}) \sum_{n=0}^{\infty}
\frac{{|\beta|}^{2n}}{n!} |n \rangle \langle n|\,.
\label{ng4.1}
\end{eqnarray}
Since ${\hat{\rho}}_{|\beta|}$ is a convex sum of Fock states, its 
 $Q$ function is a corresponding convex sum:
\begin{eqnarray}
Q^{|\beta|}(\alpha) &=& {\rm exp}[{-({|\alpha|}^2 + {|\beta|}^2)}]
\sum_{n=0}^{\infty} \frac{{|\alpha|}^{2n} {|\beta|}^{2n}}{n!n!} \nonumber \\
&=& {\rm exp}[{-({|\alpha|}^2 + {|\beta|}^2)}]
I_{0}(2|\alpha||\beta|)\,,
\label{ng4.2}
\end{eqnarray}
where $I_{0}(.)$ is the modified Bessel function of integral order zero.
The only non-zero moment 
of order $\le2$  is  
$\langle {|z|}^2 \rangle = \textrm{Tr}(\hat{\rho}_{|\bt|}
\hat{a}\hat{a}^{\dagger}) =
1+ {|\beta|}^2$. The associated  Gaussian 
probability $Q^{|\beta|}_G(\alpha)$
that has the same first and second moments is thus the 
 thermal state with average photon number 
$\langle \hat{n} \rangle = {|\beta|}^2$. As we have  shown 
earlier in Eq. (\ref{ng2.15}),
the Wehrl entropy of this Gaussian state is  
$H_{W}(\hat{\rho}_{G}^{|\beta|})=1 
+{\rm 
log}
(1+{|\beta|}^2)$. 
To compute the Wehrl entropy corresponding to the original phase-averaged
coherent state, however, we  resort to numerical evaluation. 
In Fig. (\ref{ngfig2}) we present the
non-Gaussianity of $\hat{\rho}_{|\beta|}$
as a function of $|\beta|^2$, the energy of the state.
 It is seen to be a monotone increasing function  of 
$|\beta|^2$.

\begin{figure}
\scalebox{0.7}{\includegraphics[width=10cm]{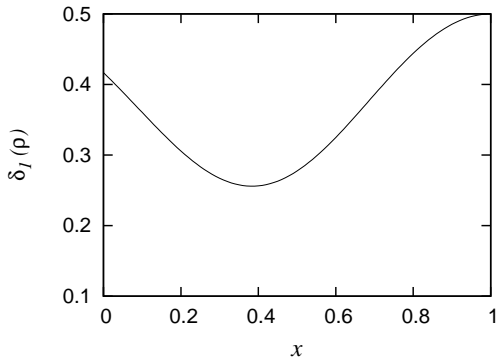}}
\caption{Variation of $\delta_1(\hat{\rho})$ as a function of the Boltzmann 
parameter $x$ for the photon-added thermal state. \label{ngfig3}}
\end{figure}

Note that the phase-averaged coherent states are classical 
since they are, by definition, convex sums of coherent states. 
Thus if a bipartite  state consisting of a phase-averaged 
coherent state in one mode and vacuum in the other is passed 
through a beam splitter, the two-mode mixed state at the output 
will remain separable (since the phase-averaged coherent 
state is classical\,\cite{solo06}),  with the same non-Gaussianity as 
the original phase-averaged coherent state.

\section{Comparison with other measures}

In this Section we compare our non-Gaussianity measure 
${\cal N}(\hat{\rho})$ with two non-Gaussianity  
measures which have been proposed recently.

\subsection{Measure based on Hilbert-Schmidt distance}
Genoni {\em et al} \,\cite{genoni07}, have proposed a non-Gaussianity 
measure 
based on the Hilbert-Schmidt distance. They define non-Gaussianity of
a state $\hat{\rho}$ as
\begin{eqnarray}
\delta_1 (\hat{\rho}) = \frac{{\rm Tr}[{(\hat{\rho} -\hat{\tau})}^2]}{2{\rm Tr}
(\hat{\rho}^2)}\,,
\label{ng5.1}
\end{eqnarray}  
where $\hat{\tau}$ is the Gaussian state with the same first and second 
moments
as $\hat{\rho}$. Let us  compare this measure with ours 
in the specific case of the PATS $\hat{\rho}_{\rm PATS}^{(m,x)}$. 
In Fig. (\ref{ngfig3}) we plot $\delta_1 (\hat{\rho}_{\rm PATS}^{(m,x)})$ as a 
function of the Boltzmann parameter $x$, for fixed value of $m= 1$. 
 It is seen that  $\delta_1 (\hat{\rho}_{\rm PATS}^{(m,x)})$, for 
$m=1$, is not a constant but varies with the temperature parameter $x$.   
This shows that
this measure of Genoni \etal does not satisfy our shape criterion.

Another interesting difference appears when one compares our measure
${\cal N}(\hat{\rho})$ with $\dl_{1}(\hat{\rho})$ in the 
case
of the photon number states $\hat{\rho}=|m\rangle \langle m|$. As we have
shown earlier\,[see Fig. (\ref{ngfig1})], our measure  monotonically 
increases with 
the photon number $m$ and tends to infinity as $m$ tends to infinity. 
In contrast, as Genoni \etal have shown and 
emphasised\,\cite{genoni07}, their measure $\dl_{1}(\hat{\rho})$ 
saturates at the  value $\half$.

\begin{figure}
\scalebox{0.7}{\includegraphics[width=10cm]{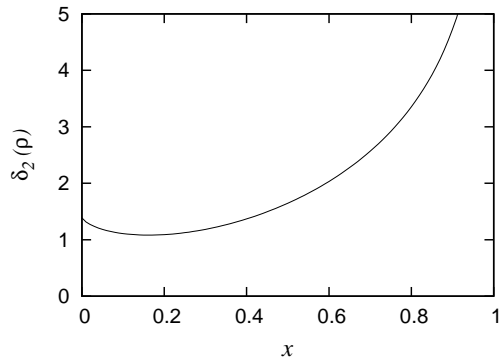}}
\caption{Variation of $\delta_2(\hat{\rho})$ as a function of the Boltzmann 
parameter $x$ for the photon-added thermal state. \label{ngfig4}}
\end{figure}

\subsection{Measure based on quantum relative entropy}
Genoni \etal\,\cite{genoni08} have proposed,  
in a subsequent paper, 
 a second measure of non-Gaussianity, this one based on quantum 
relative entropy. They define non-Gaussianity of a state $\hat{\rho}$ as 
\begin{eqnarray}
\delta_2(\hat{\rho}) = S(\hat{\tau}) - S(\hat{\rho}),
\end{eqnarray} 
where $S(\cdot)$ is the von Neumann entropy of the state in question 
and $\hat{\tau}$ is the Gaussian state 
with the same first and second moments as the given state $\hat{\rho}$. 

At first sight it would seem that $\delta_2(\hat{\rho})$ and
our measure ${\cal N}(\hat{\rho})$ are very similar, the only 
difference being that $\hat{\rho}$
is replaced by $Q(z)$ and that the trace operation in the formula for
the von Neumann entropy is replaced in our measure 
by a phase space integral. 
A closer look  reveals that this is not the case; 
$\delta_2(\hat{\rho})$
does not reduce to ${\cal N}(\hat{\rho})$ under 
 this kind of  `quantum-classical correspondence'.  
And $\delta_2(\hat{\rho})$ and ${\cal N}(\hat{\rho})$ 
turn out to be quite different entities.

A qualitative
difference between $\delta_2(\hat{\rho})$ and $N(\hat{\rho})$ becomes 
manifest
when one compares these two measures in the context 
of a pure state. 
As the von Neumann entropy of a pure state is zero, 
$\delta_2(\hat{\rho})$
reduces to $S(\hat{\tau})$, the von Neumann entropy of the Gaussian
state with the same first and second moments as $\hat{\rho}$. 
In other words 
$\delta_2(\hat{\rho})$ does not consult, in the case of pure states,  
moments or
cumulants of $\hat{\rho}$ of order higher than $2$. 
Consequently, all pure states which have the same set of first and second 
moments but differ in higher moments   
will get assigned the same non-Gaussianity $\dl_{2}(\rho)$.  
 This is  not the case with our measure ${\cal N}(\hat{\rho})$.

To bring out a second qualitative difference we check if
$\dl_{2}(\hat{\rho})$ satisfies the shape criterion. 
To this end we ask if $\dl_{2}(\hat{\rho})$ will ascribe the same amount 
of
non-Gaussianity to the PATS and the photon number state, \ie, whether 
$\dl_{2}(\rho)$ evaluated for the PATS 
$\hat{\rho}_{\rm PATS}^{(m,x)}$ 
is independent of the temperature parameter $x$.
We find that this is not the case. This is shown in 
  Fig. (\ref{ngfig4}) wherein we present 
$\dl_{2}(\hat{\rho}_{\rm PATS}^{(m,x)})$, for fixed value $m=1$, 
as a function of $x$. 
 
We conclude  this Section with a further remark. With reference to 
Figs. (\ref{ngfig3}) and (\ref{ngfig4}), while the non-Gaussianity measures 
$\dl_{1}(\hat{\rho}_{\rm PATS}^{(m,x)})$ and 
$\dl_{2}(\hat{\rho}_{\rm PATS}^{(m,x)})$, for fixed $m$, vary with 
the temperature (or scale) parameter $x$, thus failing the shape 
criterion,  
the variation is not monotone. The significance of the 
temperatures at which these measures assume their 
respective minimum values is not clear.

\section{Concluding remarks}
We have presented a measure of non-Gaussianity
of quantum states based on the $Q$ function. 
In doing so we have been guided by the fundamental
principle that any measure of non-Gaussianity 
is an attempt to make a quantitative statement 
on the departure of the shape of the $Q$ function 
from Gaussian, and the measure must therefore remain invariant 
under all transformations which do not  change the shape of the
$Q$ function. 

Uniform scaling of all the phase space coordinates at the 
level of the $Q$ function has proved to be an important 
shape preserving transformation, and our shape criterion 
demands that non-Gaussianity of the photon-added thermal states 
should be independent of temperature.

We have explored various properties our measure which meets the shape 
criterion. 
We have presented 
analytical and numerical results on the non-Gaussianity of a few
families of quantum states.  
We have also compared our measure with other measures of non-Gaussianity 
available in the literature.

Our measure ${\cal N}(\hat{\rho})$ meets the shape criterion which, 
in our opinion, should be respected by every good measure of non-Gaussianity. 
We hasten to add, 
however, that this is not the only measure that meets this criterion. 
For instance, if 
$\gamma^{(2n)}$ is an appropriate linear combination of the cumulants of 
order $2n$, and
$\gamma^{(2)}$  an appropriate linear combination of the cumulants of 
order $2$, it is clear that the ratio between  $\gamma^{(2n)}$ and the 
$n^{\rm th}$  power of $\gamma^{(2)}$ will meet this criterion, for every 
$n\ge 2$. Our choice ${\cal N}(\hat{\rho})$ has the attraction 
 of being immediately related to well-known entities like the Wehrl 
entropy and Kullback-Leibler distance. 

In the case of classical probability defined on a $2n$ dimensional 
 space ${\cal C}^n$, one would have required the non-Gaussianity 
measure to be invariant under the full Eulidean group 
consisting of tanslations and all $\rm{SO}(2n)$ rotations. 
 In the case of phase space,  $\rm{SO}(2n)$ rotations  
which fall outside the subgroup $\rm{Sp}(2n, {\cal R})\cap \rm{SO}(2n)$
 are unphysical, and hence the restriction to this subgroup of 
passive linear systems.

Our shape criterion rests on the invariance semi-group of 
$Q$ functions which is different from 
 the invariance semi-group of the Gaussian family of states --  
operations 
which map Gaussian states into Gaussians. The latter semigroup 
includes the full $\rm{Sp}(2n, {\cal R})$, and not just the intersection 
subgroup $\rm{Sp}(2n, {\cal R})\cap \rm{SO}(2n)$. It further includes a 
whole family of completely possitive maps known as Gaussian channels.

Finally, it may be noted that a scaling transformarion 
similar to the one  we have implemented on $Q$ functions cannot be 
implemented on  Wigner functions. This  follows from 
the following fact: that $W(\alpha)$ is a Wigner function does not imply 
that $\lambda^{2n}\,W(\lambda \alpha)$ is necessarily a Wigner function.

\end{document}